# Tunable Valley Polarization and Valley Orbital Magnetic Moment Hall Effect in Honeycomb Systems with Broken Inversion Symmetry

Zhigang Song[1], Ruge Quhe[1,4], Yan Li [1,2], Ji Feng[2,3], Jing Lu[1,2 *], Jinbo Yang[1,2*]

[1]State Key Laboratory for Mesoscopic Physics and School of Physics, Peking University, Beijing 100871, China

[2]Collaborative Innovation Center of Quantum Matter, Beijing, China

[3]International Center for Quantum Materials, Peking University, Beijing, China

[4]Academy for Advanced Interdisciplinary Studies, Peking University, Beijing 100871, China

Corresponding author: jbyang@pku.edu.cn, jinglu@pku.edu.cn

Abstract

In this Letter, a tunable valley polarization is investigated for honeycomb systems with broken inversion symmetry such as transition-metal dichalcogenide $MX_2$ (M = Mo, W; X = S, Se) monolayers through elliptical pumping. As compared to circular pumping, elliptical pumping is more universal and effective method to create coherent valley polarization. When two valleys of $MX_2$ monolayers are doped or polarized, a novel anomalous valley orbital magnetic moment Hall effect driven by opposite Berry curvatures at different valleys is predicted and can generate orbital magnetic moment current without the accompaniment of a charge current, opening a new avenue for exploration of valleytronics and orbitronics. Valley orbital magnetic moment Hall effect is expected to obscure spin Hall effect and is tunable under elliptical pumping.

Valleytronics have generated great interest[1-4] in the fields of condensed matter physics because it involves many novel phenomena such as valley-Hall effect[5], spin-Hall effect[6], valley magnetism[7]. The valleys refer to the conduction-band minima and valence-band maxima. In some materials, Valleys are located at ineqvalent positions of the momentum space, leading to a novel degree of freedom. Manipulation of such a degree of freedom forms the basis of valleytronics. Transition-metal dichalcogenide $MX_2$ monolayers (M = Mo, W; X = S, Se) are a perfect platform of valleytronics because of their intrinsic valley degree of freedom on honeycomb lattice and direct band gap in the regime of visible light. In order to realize 100% valley polarization, circularly polarized optical field is used to selectively excite either of the two valleys[8-10]. However, circularly polarized optical field can not tune the valley polarization. If valley polarization is tunable, physical phenomena associated with the intermediate states between the valley-unpolarized and valley-completely polarized states can be observed, for examples, valley quantum coherence[11] and some fractional-quantum-Hall states[12]. In the meantime, spin Hall effect from different valleys are enhanced, and thus valley-coupled spin Hall effect[13] becomes more apparent in a partial valley-polarized state[6].

A strong spin-orbit coupling (SOC) is required to generate valley-coupled spin Hall effect. However, SOC is not strong enough for valleytronics materials, such as graphene, silicene and silicone, to realize valley-coupled spin Hall effect. The Zeeman-type SOC splitting in the valence band maximum (CBM) of $MX_2$ monolayers is large enough (0.15 - 0.46 eV)[6, 14] to observe valley-coupled spin Hall effect with hole carriers. Unfortunately, the SOC splittings in the conduction band minimum (VBM) of $MX_2$ monolayers are too small to observe valley-coupled spin Hall effect with electron carriers. In analogy to spin Hall effect, there should be an orbital magnetic moment Hall effect,[15] which does not require SOC. Manipulation of orbital magnetic moment is the basis of orbitronics.[16] The valley-contrasting orbital magnetic moment (called valley orbital magnetic moment) in $MX_2$ monolayers consists of two parts: one is from the parent atomic orbitals, and the other is from lattice structure (valley magnetic moment). Valley orbital magnetic moment is a few times larger than spin magnetic moment in $MX_2$ monolayers. Atomic orbital magnetic moment Hall effect has been reported in *p*-doped Si bulk,[15, 16] but valley-coupled orbital magnetic moment Hall effect, especially the part from lattice structure, is rarely reported.

In this Letter, we have investigated the degree of polarization of two valleys under elliptical pumping in MX$_2$ monolayers. A valley orbital magnetic moment Hall effect driven by opposite Berry curvature at different valleys is proposed. Valley orbital magnetic moment Hall effect is thought tunable by elliptical pumping. This opens a new avenue for exploration of valleytronics and orbitronics.

In a honeycomb system with $C_{3h}$ and particle-hole symmetry, to the first order in *k*, the two-band ***k·p*** Hamiltonian near the massive Dirac cone *K* (*K'*) point without SOC can be described as follows[6]:

$$H = at(\tau k_x \sigma_x + k_y \sigma_y) + \frac{\Delta}{2} \sigma_z, \qquad (1)$$

where $\sigma_\alpha$ denotes the Pauli matrices for the two basis functions, *t* represents the effective hopping integral, *a* is the lattice constant, $\tau = \pm$ is the valley index, and *Δ* is the band gap. The pseudospin *τ* is a good quantum number near the *K* (*K'*) point, but it is strongly broken away from the *K* (*K'*) point. All of the above parameters can be obtained by fitting the first-principles band structures.

Different polarization states of a monochromatic light wave with electric field projecting in the *x-y* planes could yield time-dependent components $E_x$ and $E_y$ in cosine form as follows: $E_x = A_x \cos(\omega t)$ and $E_y = A_y \cos(\omega t + \theta)$, where *θ* is the phase retardation between $E_x$ and $E_y$, and the amplitude ratio is defined as $\gamma = \frac{E_x}{E_y}$. The coupling strength with optical fields of elliptical polarization is given by

$$P_\pm = P_x + \gamma e^{\pm i\theta} P_y, \qquad (2)$$

where $p_\alpha$ is the matrix element between the valence and conduction bands of the canonical momentum. $p_\alpha$ is given by

$$P_\alpha = m_e \langle u_c(\boldsymbol{k}) | \frac{1}{\hbar} \frac{\partial H}{\partial k_\partial} | u_v(\boldsymbol{k}) \rangle, \qquad (3)$$

where $|u_c(\boldsymbol{k})\rangle$ and $|u_v(\boldsymbol{k})\rangle$ are the periodic parts of the conduction and valence band of Bloch function, respectively, which are obtained by making the matrix in Eq. 1 diagonalized, and $m_e$ is the free electron mass. It is straightforward to derive an explicit form of $P_{\pm}$ from Eqs. 1-3. The $\boldsymbol{k}$-resolved degree of elliptical polarization is defined as

$$\eta(\boldsymbol{k}) = \frac{|p_+(\boldsymbol{k})|^2 - |p_-(\boldsymbol{k})|^2}{|p_+(\boldsymbol{k})|^2 + |p_+(\boldsymbol{k})|^2} \quad , \tag{4}$$

$\eta(\boldsymbol{k})$ is normalized by total absorption. According to Eqs.1-4, $\eta(\boldsymbol{k})$ can be expressed as

$$\eta(\boldsymbol{k}) = \frac{2\Delta\tau \sin(\theta)\gamma\sqrt{a^2t^2k^2 + \frac{\Delta}{2}(^2)}}{(a^2t^2k^2 + \frac{\Delta^2}{2})(1+\gamma^2) - (1-\gamma^2)(a^2t^2k_x^2 - a^2t^2k_y^2) - 4\gamma^2\cos(\theta)a^2t^2k_xk_y^2} \quad . \tag{5}$$

Then, the degree of elliptical polarization near the $K$ ($K'$) point is given by

$$\eta(\text{K}) = \frac{2\tau\sin(\theta)\gamma}{(1+\gamma^2)}. \tag{6}$$

The term $-(1-\gamma^2)(a^2t^2k_x^2 - a^2t^2k_y^2) - 4\gamma\cos(\theta)a^2t^2k_xk_y$ in Eq. 5 indicates that $\eta(\boldsymbol{k})$ is anisotropic. However, this term is a second-order small one and negligible in the vicinity of the $K$ ($K'$) point.

According to Eq. 6, the value of $\eta(K)$ can continuously vary from +1 to -1 as the optical helicity of the excition light changes. The helicity of the incident light can be continuously tuned by both changing $\theta$ using electronically-controlled-liquid crystal retarders[17] and adjusting the angle between polarization direction of the incident light(linearly polarized) and the fast axis of the modulator angle ($\varphi$) (see Fig. 1(a)). It is obvious that $\gamma = \tan(\varphi)$ when a quarter-wave modulation is used. Thus, $\eta(K) = \sin(2\varphi)$. Eq. 6 is established with the condition that $\Delta > 0$, namely, inversion symmetry is broken and the group of the wave vector at the band edges (the $K$ and $K'$ points) is $C_3$.

In order to futher investigate the mechanism of the valley polarization, we performed first-principles simulations of $\eta(K)$ as a function of $\theta$ and $\gamma$ based on density functional theory (DFT) with the VASP packet within the framework of the projector augmented wave (PAW) pseudopotential method using a plane-wave basis set. The exchange-correlation functional is treated with Perdew-Burke-Ernzerhof generalized gradient approximation. The cutoff energy for wave-function expansion w as set to be 350 eV. In the calculation, we only consider the direct optical excitation with $\Delta \boldsymbol{k} = 0$. There is no difference in results of $\eta(K)$ with and without SOC. Fig. 1 plots the DFT calculated values of $\eta(K)$ for MX$_2$ monolayers as a function of $\gamma$ and $\theta$. The results coincide with the analytical ones from Fig. 1 (a).

As an example, the spin-resolved degrees of elliptical polarization $\eta(\boldsymbol{k})$ for WSe$_2$ monolayer in irreducible Brillouin zone (BZ) with four different the phase retardations are shown in Fig. 2. Our DFT calculation implies that the probability of the spin-flipped inter-band transitions is 3 orders of magnitude smaller than that of spin-conservation inter-band transitions. Thus the spin-flipped inter-band transitions can be neglected. We find that $|\eta(\boldsymbol{k})| \cong |\eta(K)|$ in a large region around the $K$ ($K'$) point, because the bottom of valence bands and the top of conduction bands consist of mainly $d$-orbital character from the M atoms in a large region of the BZ. The BZ can be divided into two parts, one around the $K$ ($K'$) point, and the other around the $\Gamma$ point. The former is spin-independent, where SOC is zeeman type. The later part is spin-dependent, where SOC is Rashba type associated with the nearest-neighbor hopping. We find that the $\eta(\boldsymbol{k})$ around the $\Gamma$ point is spiral, determined by the Rashba-type spin texture[18]. It is interesting to find that $\eta(+,\uparrow) = -\eta(-,\downarrow)$ with $\boldsymbol{k}$ far from the $\Gamma$ point, where $\tau = \pm$ is a good quantum number; while $\eta(\uparrow) = -\eta(\downarrow)$ with $\boldsymbol{k}$ near the $\Gamma$ point, where $\tau$ is not a good quantum number.

When an in-plane uniaxial stress or strain is applied in MX$_2$ monolayer, the $C_3$ symmetry of MX$_2$ monolayer is broken. Valley-selective elliptical dichroism takes the place of the circular dichroism. A right-hand (left-hand) elliptically polarized photon could be selectively absorbed around the $K$ ($K'$) point, while a left-hand (right-hand) one was completely prohibited. This conclusion is similar to the recent experiment in which electronic field is used to break the $C_3$ symmetry of WSe$_2$ monolayer[2]. The calculated elliptical polarization $\eta(K)$ under different stresses or strains are shown in Fig. 1 (d). It is obvious that the maximums of $\eta(K)$ deviate from $\varphi = \dfrac{\pi}{4}$. This indicates that an elliptically polarized light is more universe than a circularly polarized light to create complete valley polarization. Therefore, elliptical dichroism could be used to detect the planar stress and strain in $MX_2$ monolayers.

In order to understand the relationship among valley magnetization, valley orbital magnetic moment Hall effect, valley Hall current, and the degree of elliptical polarization, we proceed to derive analytic expression of local chemical potential $\mu$, (see details in supplementary information)[19]

$$\mu = \frac{\Delta}{2} + \frac{eat}{\hbar}\sqrt{\pi\alpha\frac{TI}{\omega}((1+\frac{\Delta^2}{(\hbar\omega)^2})+\frac{2\Delta\tau\sin\varphi 2}{\hbar\omega}} , \qquad (7)$$

where $I$ is the intensity of the incident light, and $\alpha$ is the probability that a photon is absorbed. Local chemical potential $\mu$ as a function of the amplitude ratio $\gamma$ is plotted in Fig. 3(a).

According to particle-hole symmetry without including SOC, Valley orbital magnetizations of the valence and conduction bands are similar. The valley magnetization contributed from the lattice structure in the conduction band can be described by the Berry curvature as follows[5, 20]:

$$m_l = \frac{2e}{\hbar}\int\frac{d^2k}{(2\pi)^2}\mu\Omega(k)$$

$$= \frac{\tau e(2\mu - \Delta)}{4\pi\hbar} , \qquad (8)$$

where $\Omega$ is Berry curvature and shown in Fig 4(a). The integration is over all states below $\mu$. The valley magnetism $m_l$ is measurable. The orbital magnetization of $m_l$ as a function of $\varphi$ is plotted in Fig. 3(b), and the peaks are at $\varphi = \pm\frac{\pi}{2}$.

The valley contrasting magnetic moment originating from the lattice structure has the format: $\mu_v = \frac{\tau ea^2t^2}{\Delta\hbar} = \frac{2\tau a^2t^2 m_0}{\hbar^2\Delta}\mu_B$ in the low power limit $k \to 0$. $\mu_v$ is in close analogy with spin to the electron except that the isotropic part of the effective mass in $\mu_v$ takes place of the free electron mass $m_0$ in Bohr magneton[5]. The valley magnetic moments are in same direction in the conduction bands and value bands although they are along opposite direction at the $K$ and $K'$ points (see Fig. 4). Around the $K$ ($K'$) point of $MX_2$ monolayers, the particle-hole symmetry is not perfect, thus Berry curvatures of the conduction band minimum and the valence band maximums VBM are not absolutely opposite and the valley magnetic moments of CBM and VBM are inequivalent. Values of $\mu_v \approx 2.4\mu_B$ for the CBM and $\mu_v \approx 3.6\mu_B$ for the VBM are obtained in WSe$_2$ monolayer.

In addition to the lattice structure contribution, the parent atomic orbital magnetic moment also contributes to the orbital magnetic moment. The CBM is mainly composed of $d_{z^2}$-orbitals with $m = 0$, while the VBM is mainly of $d_{x^2-y^2} + id_{xy}$-orbitals with an atomic orbital magnetic moment of $2\mu_B$ round the $K$ point and $d_{x^2-y^2} - id_{xy}$-orbitals with an atomic orbital magnetic moment of $-2\mu_B$ around the $K'$ point.[10] The valley magnetic moments are in the same direction in the conduction and valence bands although they are along opposite direction at the $K$ and $K'$ points. Atomic orbital magnetic moment is also about 2 times of spin moment, and its direction is the same as one of valley magnetic moment[21]. Therefore, the valley orbital magnetic moment of CBM is about 2.4 times of spin magnetic moment, and that of VBM is 5.6 times of spin magnetic moment in monolayer WSe$_2$. The valley orbital magnetic moments of the two conduction (valence) bands due to SOC splitting are parallel, while the spin moments are antiparallel.

It was well established that an electron will acquire an anomalous transverse velocity when an in-plane electric field is applied[5], which is proportional to the Berry curvature. Thus the carriers from different valleys carry opposite valley orbital magnetic moments and flow to opposite transverse edges in the presence of an in-plane electric field, resulting in an anomalous orbital magnetic moment Hall effect with valley Hall effect in close analogy. Valley orbital magnetic moment current ($J$) is defined as the average of the valley orbital magnetic moment per electron ($\bar{m}_\tau$) times the velocity operator $J_\tau = \bar{m}_\tau \vec{v}$, $\vec{J} = \sum_\tau \sigma_\tau \frac{\bar{m}_\tau}{e} \hat{z} \times \vec{E}$, where $\vec{E}$ is electric field, $\hat{z}$ the unit vector in the vertical direction, and $\sigma_\tau$ the valley Hall conductivity and given by [6]

$$\sigma_\tau = \frac{2e^2}{\hbar} \int \frac{d^2\mathbf{k}}{(2\pi)^2} f(\mathbf{k},\mu)\Omega(\mathbf{k}),$$

$$= \frac{\tau e^2 (2\mu - \Delta)}{4\pi\hbar\mu} \quad , \tag{9}$$

where the integration is over all states below $\mu$. There are three types of valley orbital magnetic moment Hall effect: electron-doped, hole-doped and electron-hole pair type respectively (see Fig. 5). The former two are accompanied with no charge current. $\mu$ is a function of $\theta$ and $\varphi$, and thus $\sigma_\tau$ can be adjusted by the helicity of incident light in the last case. If the edge states of samples are insulating, there is an accumulation of carriers with opposite orbital magnetic moment and Berry phase on opposite sides, which can be detected by Kerr or Faraday effects. If the edge states of samples are metallic, there may be one-dimensional channel for valley orbital magnetic moment. Although orbital magnetic moment current is not directly measurable, magnetism current can induce an electric field, which is an indirect signal to detect magnetism current.

If the average valley orbital magnetic moment $\bar{m}_\tau$ is replaced by electric charge, the common valley charge current is obtained:

$$j_x = \frac{e^2 (\mu_K - \mu_{K'})}{4\pi\hbar\mu_K\mu_{K'}} E_y \quad . \tag{10}$$

Similar to the valley Hall conductivity, the valley charge current is tunable by the helicity of incident light Fig. 3 (c) and (d) show the valley Hall conductivity of the conduction bands ($\sigma_\tau$) and the valley charge current $j_x$ ( $j \propto J$ ) vs. angle $\varphi$, respectively. The curve of the valley charge current dependent of $\varphi$ shows a sine-like format. Recently, valley Hall effect was successfully observed[1]. The observed dependence and period of the valley electric current on $\varphi$ are in agreement with our theoretical results (see Fig. 3(d)).

$MX_2$ monolayers are different from graphene mainly in two folds. One is that inversion symmetry is explicitly broken in $MX_2$ monolayers, and the other is that the band gaps of $MX_2$ monolayers are in the visible frequency range. Both features are critical to realize valley orbital magnetic moment Hall effect. A large band gap corresponds to a long lifetime of electron-hole pairs and broad distribution region of Berry curvature in momentum space. As a result, the number of the carriers carrying nonzero valley orbital magnetic moment is large, resulting in a large orbital magnetization according to Eq. 8. In addition, since the two valleys in momentum space is widely separated, the intervalley scattering is strongly suppressed in $MX_2$ monolayers and graphene[22]. The combination of long lifetime of electron-hole pairs and strongly suppressed intervalley scattering is expected to get a long lifetime of the carriers. Actually, long lifetime and high-density carriers have been realized only in $MX_2$ monolayers by circular pumping at the corners of the Brillouin zone[8, 23]. Hence, $MX_2$ monolayers are perfect systems to realize and observe valley orbital magnetic moment Hall effect. In contrast, the band gap of graphene opened by inversion asymmetry is very small (< 0.3 eV and in the infrared region)[24, 25], and Berry curvature is localized in the vicinity of the $K(K')$ point. Therefore, the number of carriers carrying nonzero valley orbital magnetic moment is small, resulting in a small orbital magnetization according to Eq. 8, although the valley magnetic moment is very large. Besides, the lifetime of electron-hole pairs at massive Dirac cones is very short in the graphene due to the small band gap. Taken together, it is difficult to realize and observe valley orbital magnetic moment Hall effect in the graphene.

In summary, a concise analytical expression of $\eta(K)$ dependent of the phase retardation between the two components of the electric field $\theta$ and the amplitude ratio $\gamma$ is dervied. It is found that elliptical pumping is a more general and effective method to create coherent valley polarization. There is a novel valley orbital magnetic moment Hall effect driven by opposite Berry curvature at different valleys, which is expected to mask spin Hall effect. The valley orbital magnetic moment Hall effect generates orbital magnetic moment current without the accompaniment of a charge current. Besides valley orbital magnetic moment Hall effect is tunable under elliptical pumping

**ACKNOWLEDGMENT**

This work was supported by the MOST Project of China (Nos. 2010CB833104 and 2013CB932604), the National Natural Science Foundation of China (Nos. 51371009, 50971003, 51171001, and 11274016).

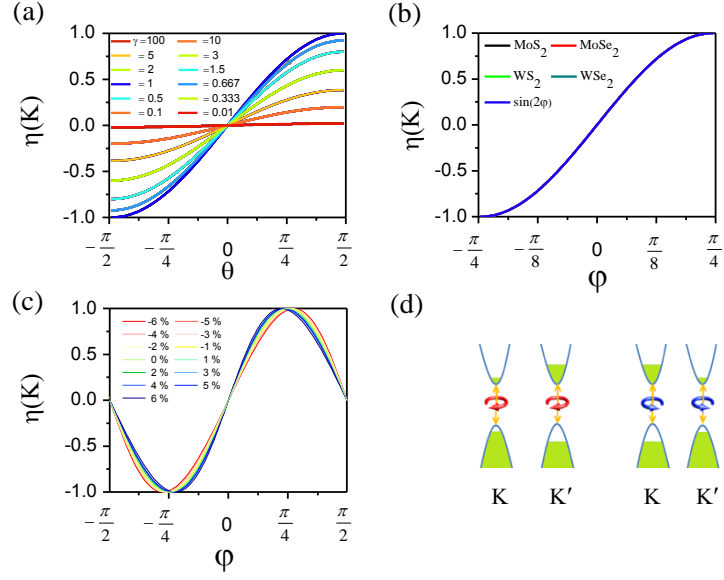

FIG. 1 (color online) (a) Degree of elliptical polarization at the *K* point *η*(*K*) as a function of the phase retardation *θ* between the two components of electric field and the amplitude ratio *γ* calculated from Eq. 6 and the DFT method in MX$_2$ monolayers. The deviation between analytic and numerical results is very small. (b) *η*(*K*) as a function of rotation angle of the quarter wave plate *φ*. SOC has no influence on valley polarization around the *K* (*K'*) point. Each visible curve consists of five curves respectively fitted from Eq. 6 and calculated from MX$_2$ monolayers. (c) *η*(*K*) of WSe$_2$ monolayers under different uniaxial stresses or strains as a function of *φ* calculated from the DFT method. (d) Coherent valley polarization is created by right-(left-)hand elliptical pumping.

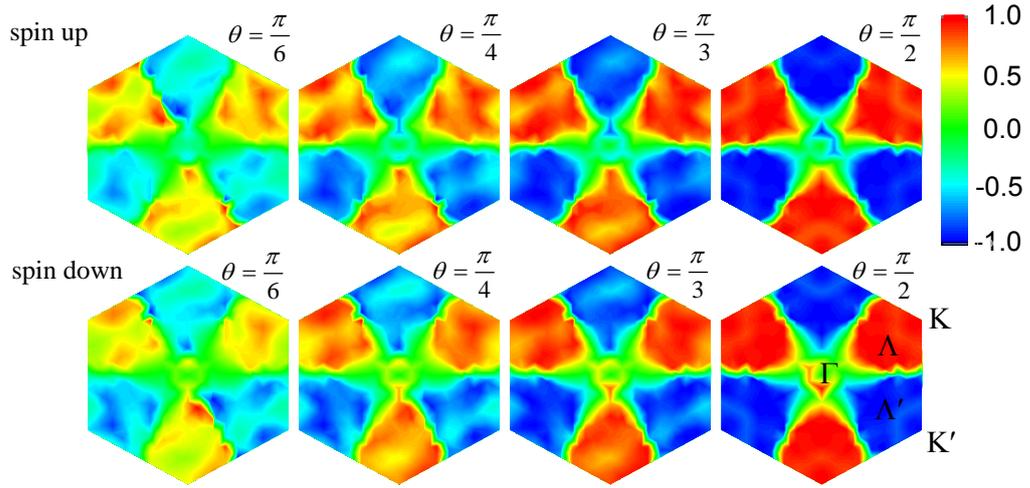

FIG. 2 (color online) Spin-resolved degree of elliptical polarization $\eta(\boldsymbol{k})$ in irreducible Brillouin zone in WSe$_2$ monolayer for the phase retardation between the two components of electric field with $\theta = \pi/6$, $\pi/4$, $\pi/3$, and $\pi/2$, respectively, calculated by the DFT method. The magnitude ratio is $\gamma = 1$.

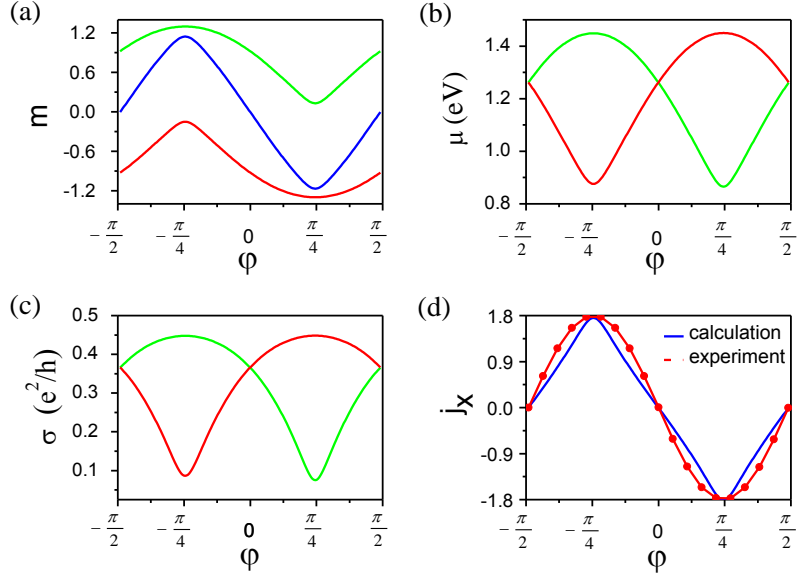

FIG. 3 (color online) (a) Local chemical potential, (b) valley orbital magnetization, (c) valley Hall conductivity, and (d) valley charge current in $MX_2$ monolayers as a function of the rotation angle of the quarter wave plate $\varphi$. The red curve corresponds to the $K$ point, green to the $K'$ point, and blue to the total. All these data are calculated from the analytic method. The experimental valley Hall current[1] as a function of $\varphi$ is given as a comparison.

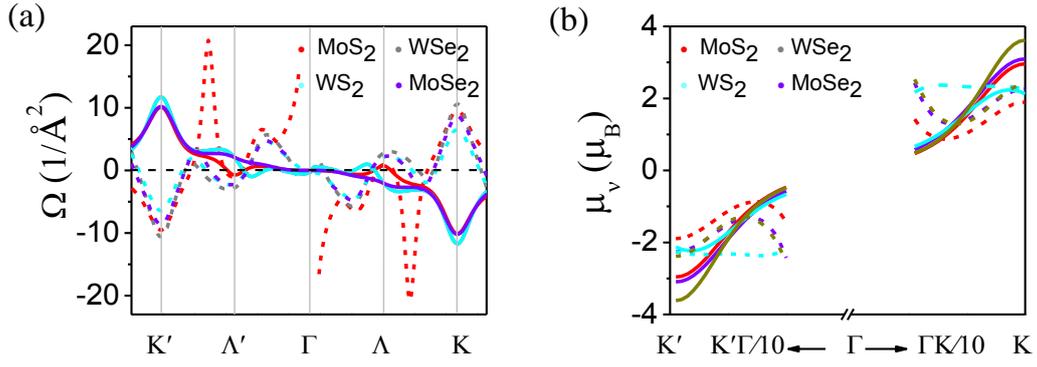

FIG. 4 (color online) (a) Berry curvature $\Omega(\boldsymbol{k})$ and (b) valley magnetic moment $\mu(\boldsymbol{k})$ of MX$_2$ monolayers calculated from the DFT method without including SOC. The solid and dashed curves represent values of the valence and conduction bands, respectively. The Berry curvatures of the conduction bands are cut off near the $\Gamma$ point, where the values of Berry curvature are exceptionally large as a result of the band degeneracy.

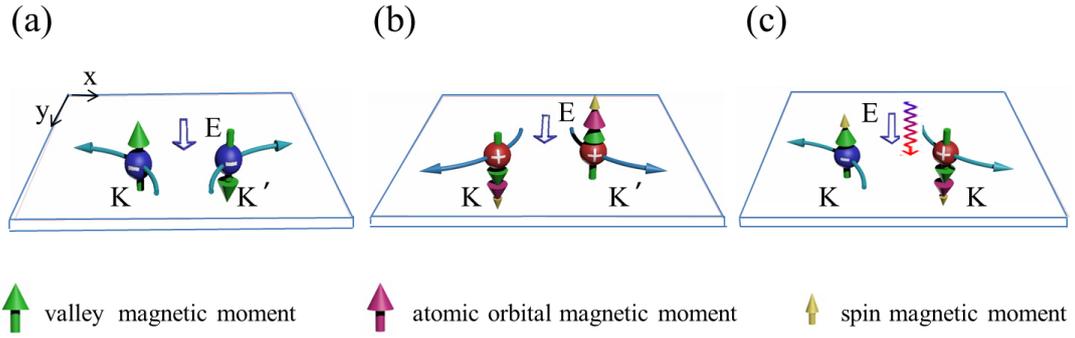

FIG. 5 (color online) (a) Valley orbital magnetic moment Hall effects of hole-doped systems, (b) Valley orbital magnetic moment Hall effects of electron-doped systems, (c) Valley orbital magnetic moment Hall effect of electrons and holes excited by circularly polarized optical field. The blue spheres are electrons, while red ones are holes. Light blue arrows indicate movement of electrons or holes. Empty arrows indicate directions of electric fields. The helical line is light beam with left or right charity.